\PassOptionsToPackage{greek,english}{babel}
\PassOptionsToPackage{protrusion=true,expansion=true}{microtype}
\PassOptionsToPackage{pdftex,final,breaklinks,bookmarks=false,bookmarksnumbered,plainpages=false,pdfpagelabels,pdfborder={0 0 0},hyperindex}{hyperref}
\PassOptionsToPackage{table,svgnames,x11names}{xcolor} 
\PassOptionsToPackage{scaled=.8}{beramono} 
\documentclass[fleqn, 5p,times,letterpaper, preprint]{elsarticle} 

\usepackage{microtype}
\usepackage[utf8]{inputenc}
\usepackage[T1]{fontenc}
\usepackage[english]{babel}

\usepackage{textcomp}
\usepackage{paralist}
\usepackage{enumerate}
\usepackage{pifont}
\usepackage{graphicx}
\usepackage{subcaption}
\usepackage{url}
\usepackage{ifthen}
\usepackage{xspace}
\usepackage{multicol}

\usepackage[scaled=.8]{beramono}
\usepackage[noteubner]{showcode}

\setmintedinline{breaklines}

\def\defaultfontsize{\small}
\def\defaultshellfontsize{\small}
\def\defaultreplfontsize{\small}

\usepackage{colortbl}
\usepackage{array}
\usepackage{multirow}
\usepackage{booktabs}
\usepackage{textcomp}

\newcommand{\tool}[1]{\textsf{#1}}

\usepackage{csvsimple}
\usepackage{hyperref}
\usepackage{algorithm}
\usepackage{algpseudocode}
\usepackage{orcidlink}
\usepackage{tikz}
\usetikzlibrary{calc, positioning, matrix, shapes, shapes.callouts, shapes.misc, shapes.symbols, shapes.geometric, shapes.multipart, fit, shadows, decorations, shapes.arrows, arrows, trees, mindmap, intersections, backgrounds, positioning, decorations.shapes, decorations.pathreplacing, decorations.pathmorphing, decorations.markings}

\usepackage[inline]{enumitem}

\newcommand{\devnull}[1]{}

\tikzstyle{every picture}+=[remember picture]
\tikzstyle{na} = [baseline=-2.5pt]

\tikzstyle{block} = [rectangle, draw, fill=white!20, text width=12em, text centered, rounded corners, minimum height=4em]
\tikzstyle{line} = [draw, -latex']

\renewcommand{\raggedright}{\leftskip=0pt \rightskip=0pt plus 0cm}

\newtcbox{\inlinebox}[1][]{enhanced,
 box align=base,
 nobeforeafter,
 colback=black!10,
 colframe=black,
 size=small,
 left=0pt,
 right=0pt,
 boxsep=2pt,
 #1}

\newcommand{\sect}[1]{\S\ref{#1}}

\begin{document}

\begin{frontmatter}

   \title{Generalized Software Product Line Extraction}

   \author[unimi]{Federico Bruzzone\,\orcidlink{0009-0004-6086-8810}} 
   \ead{federico.bruzzone@unimi.it}
   \author[unimi]{Walter Cazzola\,\orcidlink{0000-0002-4652-8113}\corref{cor1}} 
   \ead{cazzola@di.unimi.it}
   \author[unimi]{Luca Favalli\, \orcidlink{0000-0001-7452-2440}} 
   \ead{luca.favalli@unimi.it}
   \affiliation[unimi]{
       organization={Università degli Studi di Milano, Computer Science Department},
       city={Milan},
       country={Italy}
   }
   \cortext[cor1]{Corresponding author.}
   \begin{abstract}
   Software product line (SPL) engineering has been successfully applied to software development by obtaining software systems as compositions of modular features. Existing approaches to SPL engineering, however, are typically bound to a specific technological space (such as, a programming language and a composer) and \textit{integrated development environment} (IDE), and rely on extraction mechanisms that make strong assumptions on the underlying technological space. This tight coupling hinders reuse, evolution, and adoption of heterogeneous development environments.
We propose a general, workbench-agnostic protocol for extracting feature models from existing software artifacts and for configuring and deriving software products. The protocol follows a bottom-up approach based on lightweight dependency units called ``atoms'', and organizes the extraction and configuration process around an SPL server (workbench-independent) and an SPL client with a workbench-specific backend and a generic frontend. The protocol makes few assumptions on the underlying software artifacts and is therefore applicable to varied SPLs.
The applicability of this approach is presented through a prototypical implementation of the architecture in which several subsystems interact and can be swapped freely without affecting the others.
In particular, we focus on the application of such a protocol in the context of language product lines (LPLs), demonstrating its applicability to concrete scenarios while preserving workbench-agnosticism.
From bottom to top, the implementation comprises: \textsf{Neverlang} language artifacts, a \tool{Java} SPL client backend, an agnostic and reusable SPL server written in \tool{Go} and \tool{Prolog}, and a \tool{JavaScript} SPL client frontend.

   \end{abstract}

\end{frontmatter}

\section{Introduction}\label{sect:introduction}

\noindent\textbf{Context.}\quad
Modern software systems are expected to meet the demands of a continuously expanding user base. Given the inherent diversity in user requirements, software must be both highly customizable and easily reconfigurable. To address this, researchers and practitioners have increasingly focused on software product lines (SPLs), which draw inspiration from industrial product line practices to support the development of highly variable systems.

Among software systems, programming languages constitute a particularly interesting case: despite the large variety of programming languages in existence, the need for novel programming techniques and innovative language constructs drives the creation of new programming languages, both general-purpose (GPL) and domain-specific (DSL).
Application domains are varied and include concurrency, querying, and hardware design. Despite sharing many common language features, these languages are typically implemented using monolithic compilers or interpreters. This is inefficient, as it prevents language developers from reusing existing features from other languages. Moreover, extending languages is difficult, since implementations are deeply tangled, leading to highly requested features such as pattern matching in \tool{Java} taking years to be fully implemented into the language. Similarly, \inlinerust{specialization} in \tool{Rust} has remained an unstable feature for years. It is still not available in the stable release of the language due to the complexity of its implementation and potentially type system unsoundness issues~\cite{Bruzzone26d-preprint, Turon17, Matsakis18}, further illustrating how monolithic compiler architectures hinder the incremental adoption of new language constructs.

The SPL approach suits the development of programming languages well, as it addresses this limitation by building compilers and interpreters as compositions of modular language features: the compiler or interpreter emerges as the result of the composition process among features selected from a feature model. SPLs whose products are programming languages are dubbed \textit{language product lines} (LPL)~\cite{Cazzola13g, Cazzola16i}.

\noindent\textbf{Problem Statement.}\quad
Two main approaches exist for constructing SPLs~\cite{Cazzola16i}. In the top-down approach, a feature model is designed first, then features are mapped to software artifacts, and finally software variants are generated through a configuration process. Manually designing is time-consuming, then this approach is best suited to creating SPLs from scratch, whereas changing the feature model at a later point is ill-advised, as changes to the feature model may affect both the feature implementations and the final products.
In contrast, the bottom-up approach reduces the effort required for variability analysis and feature modeling by extracting feature models from existing software artifacts. Compared to the top-down approach, it trades control for easier scalability and maintainability, as it intuitively supports the incremental addition of new components and features while automatically updating both the feature model and its mappings without human intervention.

A major challenge in feature model extraction is the lack of general protocols applicable across different technological spaces. Existing tools for feature model extraction are tightly coupled to specific platforms.
Let us focus on LPLs as an example.
\tool{AiDE}~\cite{Cazzola20} is designed to work solely with \tool{Neverlang}, on the \tool{Eclipse} platform and powered by \tool{FeatureIDE}~\cite{Thum14}. Similarly, \tool{Spoofax}~\cite{Wachsmuth14} provides its own wizard for creating \tool{Eclipse} projects with skeletal language definitions and grammar extensions.
The state of the art in LPL engineering is therefore in direct contrast with the current language design trend, which is shifting towards letting developers choose their own IDE and related tools. This is evident when looking at contributions such as the \textit{language server protocol} (LSP)\footnote{\url{https://microsoft.github.io/language-server-protocol/}} which is considered the \textit{de facto} standard in IDE development.

A similar issue affects SPLs in general: the tooling for \tool{AHEAD}~\cite{Batory04}, \tool{FeatureHouse}~\cite{Apel09} and even \tool{AspectJ}~\cite{Kiczales01} are tied to (or at least intended for) a specific IDE.

\noindent\textbf{Contribution.}\quad
The contribution of this work is a generalized protocol enabling feature model extraction and supporting configuration processes. To achieve this, we made virtually no assumptions about how software features are represented, how composition is performed, or which IDE is used. This led us to focus on the fundamental structure of SPLs, independent of constraints imposed by specific language workbenches.

To validate the approach, we created a prototypical implementation that extracts a LPL from \tool{Neverlang} language artifacts, comprised of a server written in \tool{Go} and \tool{Prolog}, a client backend written in \tool{Java} and a client frontend written in \tool{Javascript} using \tool{Cytoscape}.\footnote{\url{https://cytoscape.org/}}
Given a \tool{Neverlang} codebase, the user can interact with the generated feature model and derive a configuration through the graphical user interface. Although the implementation itself is not intended as a full-fledged contribution, but rather as proof of concept, we assess the potential of the protocol in terms of lines of code saved in case the server implementation, the frontend, or both were to be reused in a different technological space.

\noindent\textbf{Structure.}\quad
The remainder of this paper is structured as follows: \sect{sect:background} presents any background information needed to understand our contribution; \sect{sect:example} introduces \tool{LogLang} a \tool{Neverlang}-based language product line that will serve as both a running example and as an evaluation case study; \sect{sect:contribution} describes the protocol in detail; \sect{sect:casestudy} presents a prototypical implementation of the protocol and its usage, while providing a rough estimate of the associated implementation effort. Finally, in \sect{sect:related-work} and \sect{sect:conclusion} we present any related works and draw our conclusions on this work.

\section{Background}\label{sect:background}
This section introduces the foundational concepts underlying our work, namely software product lines, feature modeling, language product lines, and language workbenches.
While our contribution is not tied to a specific technological space, we also briefly present \tool{Neverlang} and \tool{AiDE}, as they will be relevant to discuss parts of the protocol and of the proof-of-concept implementation.

\subsection{Software product lines and feature modeling}\label{ssect:fm}
Variability-rich software system development builds on principles from SPL engineering and feature-oriented programming. An SPL is a family of related software products whose commonalities and variabilities are expressed in terms of \textit{features}~\cite{Apel13}. SPL engineering combines \textit{domain engineering}, which defines reusable software artifacts, with \textit{application engineering}, which derives concrete products by selecting and composing features.
A central activity in SPL engineering is \textit{feature modeling}, introduced by the FODA method~\cite{Kang90}. A \textit{feature model} captures the variability of a system by describing features and their dependencies. A product is derived by selecting a valid subset of features, called a \textit{configuration}. Features belonging to a configuration are \textit{active}, while all others are \textit{inactive}. Configuration validity is determined by dependencies expressed either implicitly---through mandatory, optional, alternative, and grouped features, as well as parent---child relations---or explicitly via cross-tree constraints formulated as boolean expressions.
Both implicit and explicit dependencies may introduce anomalies such as dead features, false-optional features, and \textit{atomic sets} of features that always co-occur. Detecting and preventing such anomalies through static analysis is an active research area, encompassing both structural~\cite{Benavides10} and behavioral~\cite{TerBeek19} approaches.
When designing a generalized protocol for LPL extraction, a generalized protocol must be able to capture each of these elements.

\subsection{Language product lines}\label{ssect:lpl}
Applying SPL principles to language development leads to LPLs~\cite{Cazzola15f}, in which interpreters and compilers are derived from combinations of language features rather than implemented as monolithic systems.
LPL engineering facilitates language reuse and variability management, allowing languages to be tailored for specific purposes such as security (\textit{e.g.}, \tool{Java Card}~\cite{Chen00}), education~\cite{Bettini15,Cazzola16}, or the incremental extension of existing languages with new constructs, such as type-checked SQL queries~\cite{Erdweg11}. Variants of state-machine languages, for instance, have been successfully modeled as a single language family (e.g.,~\cite{Tratt08,Crane05,Cazzola13g}).
LPLs can be constructed following either a \textit{top-down} or a \textit{bottom-up} approach~\cite{Cazzola16i}. The top-down approach starts from a feature model derived through domain analysis~\cite{Pohl05,Apel13}, after which features are mapped to language artifacts and configured into variants. The bottom-up approach, in contrast, derives the feature model automatically from existing language artifacts. This latter approach aligns with extractive and reactive SPL adoption techniques~\cite{Krueger01} and is the focus of this work.

\subsection{Neverlang and AiDE}\label{sec:preliminaries:ssec:neverlang}
\tool{Neverlang}~\cite{Cazzola12c,Cazzola15c} is a language workbench that supports the modular development of programming languages. Language features are encapsulated in independent components called \textit{slices}, which can be compiled, tested, and reused across different language implementations. Slices are built from \textit{modules} defining reference syntax and \textit{roles} implementing semantic actions according to syntax-directed translation principles~\cite{Aho86}.
Dependencies between modules arise from shared attributes accessed or produced by semantic actions, enabling an explicit representation of feature interactions.
Ensuring the soundness of such dependencies under separate compilation---\textit{i.e.}, guaranteeing that no attribute is accessed before being defined---has been recently addressed through \tool{nlgcheck}~\cite{Bruzzone26c-preprint}, a static analysis tool based on data-flow analysis that detects potential runtime errors at compile time while preserving the modularity benefits of separate compilation.
Modules are composed into slices, and slices are further composed to produce a complete language implementation. Grammar composition, renaming mechanisms, and role sequencing enable flexible language assembly while preserving modularity.

\tool{AiDE}~\cite{Cazzola13g,Cazzola14e} extends \tool{Neverlang} with variability management capabilities for LPL engineering. It extracts feature and dependency information from language modules and synthesizes a bottom-up feature model~\cite{Cazzola20}. Its graphical interface allows developers to explore the generated model, configure variants, resolve dependencies, and test viable language instances.
\tool{Neverlang} and \tool{AiDE} provide a suitable case study: the former supports LPL engineering while the latter, though tied to a specific technology, uses a feature model extraction algorithm~\cite{Cazzola13g,Cazzola14e,Cazzola15f} that can be generalized for the purpose of this work.

\subsection{PROgrammation en LOGique}
Colmerauer \textit{et al.}~\cite{Colmerauer96} designed \tool{Prolog} in the 1970s around the idea that programs can be expressed as logical descriptions of knowledge. It relies on a subset of first-order predicate logic and builds on seminal work in automated theorem proving~\cite{Herbrand68, Robinson65}.
Although conceived for natural language processing, its conciseness and expressive power led to adoption in domains such as formal logic, knowledge representation and reasoning, and database programming~\cite{Nugues14}.

In \tool{Prolog}, \textit{facts} are statements describing properties of objects or relationships between objects. For example, to encode that Ulysses, Penelope, Telemachus, Achilles, and others are characters appearing in Homer’s \textit{Iliad} and \textit{Odyssey}, we can declare a set of facts as shown below.

\showprolog*{characters.pl}

A collection of facts, and later of rules, constitutes a \textit{database} encoding the knowledge of a given domain in logical form. By adding further facts, we can express additional properties like gender
or relationships like parenthood.

\showprolog*{charactersSex.pl}
\showprolog*{charactersParents.pl}

Given a database, a \textit{query} is a request to retrieve or verify information from the database.
Prolog answers \inlineprolog{Yes} if the query can be proven from the available facts and rules, and \inlineprolog{No} otherwise. For instance, the question ``Is Ulysses male?'' corresponds to the following query. 

\showbashcon{isUlyssesMale}

Whereas the same question posed for Helen yields a negative answer.

\showbashcon{isHelenMale}

Variables---syntactically beginning with an uppercase letter or an underscore---can unify facts within the database.
During query evaluation, \tool{Prolog} searches for facts or rules that unify with the query and substitutes variables accordingly~\cite{Nugues14}. This mechanism enables queries such as ``Who is Telemachus’ father?''. Multiple answers may exist depending on the query. 

\showbashcon{telemachusFather}
\showbashcon{odisseyCharacters}

Goals in a conjunctive query may share variables, thereby constraining multiple predicates to refer to the same value. For example, the question ``Who is a father in the \textit{Odyssey}?'' can be expressed as a conjunction of the goals \inlineprolog{character(X, odyssey)} and \inlineprolog{father(X, Z)}, where the variable \inlineprolog{X} is shared: 

\showbashcon{fatherInOdissey}

\textit{Rules} allow the derivation of new properties or relations from existing ones. For instance, the predicate \inlineprolog{son(X, Y)} can be defined as holding when \inlineprolog{X} is male and \inlineprolog{Y} is either the father or the mother of \inlineprolog{X}. 
Rules can be queried just as facts.

\showprolog*{son.pl}
\showbashcon{telemachusIsSon}

Although of course other options are viable, \tool{Prolog} is well suited for the requirements of this work. It can be used to represent features and their dependencies in a software product line. Please refer to \sect{sect:casestudy} for more details.

\begin{figure*}[ht]
    \centering
    \includegraphics[width=.9\linewidth]{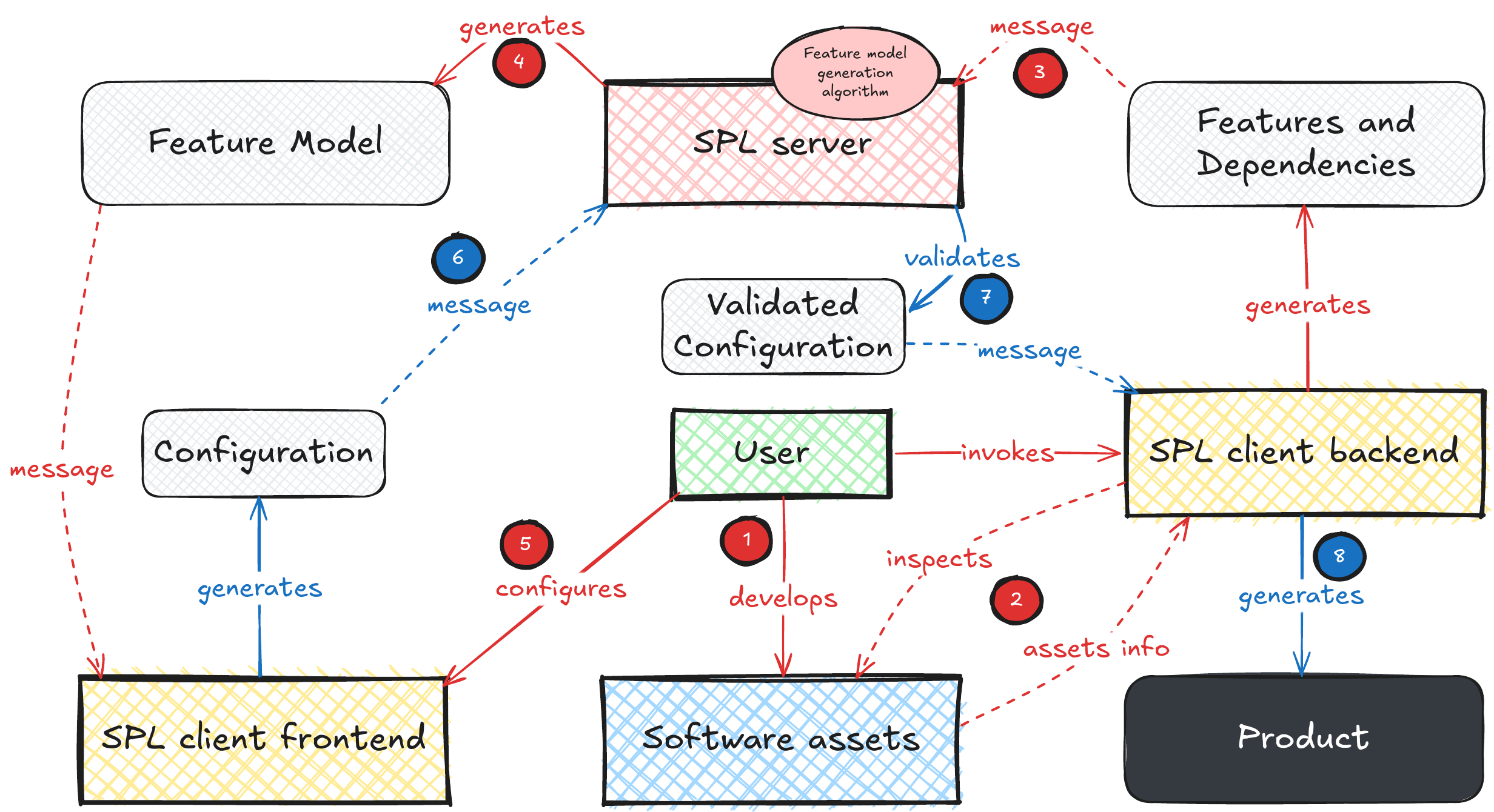}
    \caption{High-level overview of the protocol, including its actors and its steps.}%
    \label{fig:overview}
\end{figure*}

\begin{listing}[t]
    \centering
    \showneverlang*{Backup.nl}
    \caption{Excerpt of the \tool{LogLang} source code.}\label{lst:loglangsrc}
\end{listing}

\begin{figure}[t]
    \centering
    \includegraphics[width=\linewidth]{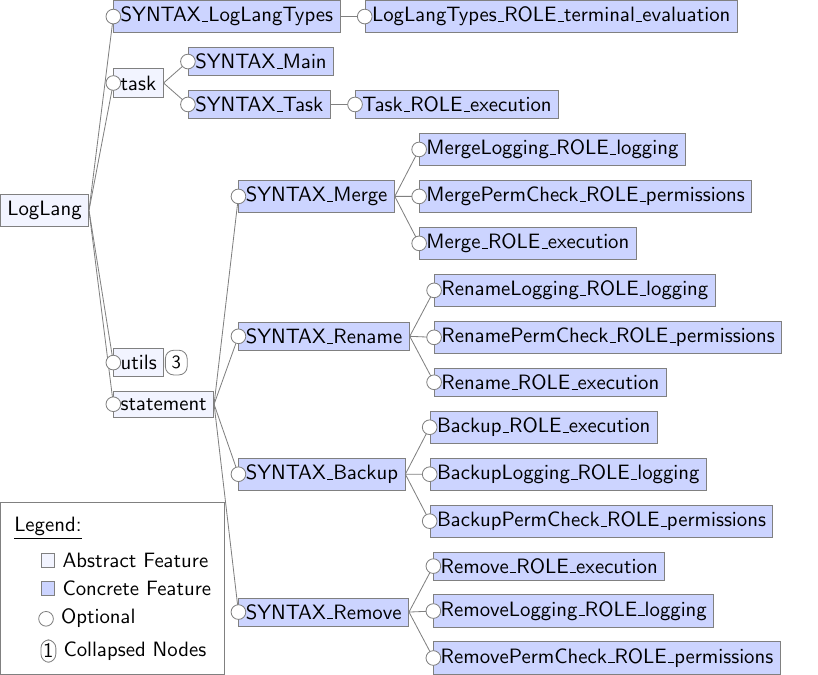}
    \caption{Feature model of the \tool{LogLang} LPL.}\label{fig:loglangfm}
\end{figure}

\section{Running Example}\label{sect:example}

\noindent\textbf{LogLang.}\quad
While our contribution does not make any assumptions with regard to the technological space being used to develop a SPL, we will present the feature model extraction protocol and our evaluation case stude in terms of its application with LPLs written in \tool{Neverlang}. We will specifically focus on the \tool{LogLang} LPL, a family of languages for the definition of log rotating tasks, similar to the \tool{logrotate} \tool{UNIX} utility.

The \texttt{Backup} module shown in Listing~\ref{lst:loglangsrc} declares a syntax for the backup task (lines~2-6), comprised of two productions. On line 5, the grammar excerpt can be tagged by the developer to declare specific implementation concerns. We will discuss how these tags are relevant in \sect{sect:contribution}.
Each role, introduced by the keyword role (line 7), defines part of the language semantics.
The specifics of how the semantics are implemented are not relevant to this work, but for easier understanding, it suffices to say that each role can be mapped to a production during the composition phase, then the evaluation is driven by passing attributes along the parse tree, according to the \textit{syntax-directed translation} technique~\cite{Aho86}.

The composition of syntax and semantic assets into a language is twofold:
\begin{enumerate*} 
   \item between modules, which yields slices, and
   \item between slices, which yields a language implementation.
\end{enumerate*}
The \texttt{BackupSlice} slice (lines~14-18) composes syntactic assets (in this case the reference syntax from the \texttt{Backup} module, line~15) and semantics assets (from two separate roles of two different modules, lines~16-17). Finally, the language descriptor (lines~19-24) composes several language features into a complete language interpreter or a compiler (lines~20-21), including other information such as parse tree visits (line 23) and relevant data structures (line 22).

Fig.~\ref{fig:loglangfm} shows an exemplary feature model for the \tool{LogLang} LPL. It contains all the language features that can be composed into a complete language specification, as well as their semantics. The features in the model comprise syntactic and semantic assets. Therefore, an SPL extraction mechanism must identify all syntactic and semantic assets from \tool{Neverlang} modules, then, upon performing the composition of a configuration into a language, the composer should be able to generate all the relevant slices (lines 14-18) and the languages unit (lines 19-24). Of course, this may vary depending on the technological space, but in \sect{sect:contribution} we will show how this can be achieved without making any assumptions in this regard.


\section{Feature Model Extraction Protocol}\label{sect:contribution}
This section details a general protocol designed to extract a feature model from a set of software assets. The goal is to make minimal assumptions about the technological space in use, so that the same protocol can be reused.

\subsection{Overview}\label{sect:contribution:overview}
Fig.~\ref{fig:overview} summarizes the most important steps of the feature model extraction protocol, as well as the additional steps that lead to the configuration, its validation and product derivation.
The process is akin to the human circulatory system, as the information is first sent across and manipulated by several components, from the core---the client backend---all the way back to the peripheries---the client frontend---and then all the way back. The way forward---highlighted in red in Fig.~\ref{fig:overview} and comprised of steps \ding{182} to \ding{186}---involves the extraction of features from the software assets, the creation of a feature model and the configuration process, whereas the way back---highlighted in blue and comprised of steps \ding{187} to \ding{189}---involves the validation of a configuration and the product derivation.

In the following, let us introduce the actors involved in the process and then detail each step of their interaction.

\subsection{Actors}\label{sect:contribution:actors}
The protocol involves the following actors.
All actors are highlighted by boxes with sharp corners and in a different color. Other entities are represented by grey boxes with rounded corners.
The final product is represented by a black box.
Please refer to \sect{sect:contribution:protocol} for more details on the behavior of each actor.

The \inlinebox[colback=green!10]{User} is the actor in charge of developing the language assets and of interacting with the feature model through a frontend application to derive a product configuration. Depending on the development process and on the project scale, the user may be a single entity or a team of multiple entities, such as language developer, language deployer and language user~\cite{Cazzola21b}.

The \inlinebox[colback=blue!20]{Software assets} are the concrete software artifacts of the product line. With regards to our running example, it comprises the linguistic assets, including both syntax and semantics, that may be part of a language product---\emph{i.e.} the reference syntax and roles from \tool{Neverlang} modules. Instead, slices and languages are our expected output (the result of the composition) and therefore are not considered part of the software assets.

The \inlinebox[colback=yellow!10]{SPL client} is the software that allows the user to interact with the services offered by the server. It behaves as an adapter that translates any concepts specific to a technological space (\tool{Neverlang} in this case) into generic SPL concepts that can be understood by the server and vice-versa. The client is comprised of two pieces of software:
\begin{compactitem}
    \item the backend is a piece of software that may be integrated within the IDE as a plugin or used in a standalone fashion; its two tasks are supervising the project structure to collect any software assets to be fed to the server
    and translating a validated configuration into a product;
    \item the frontend is an application the user can interact with to inspect the feature model, eventually deriving a product configuration; it translates the feature model metadata from the server into a (optionally) graphical and interactive representation and it sends the active features selected by the user back to the server.
\end{compactitem}
Going back to the example, the client backend should collect all relevant pieces of information from \tool{Neverlang} modules, ignoring slices and languages during the way forward, then generate the correct slice and language units depending on the configuration it received from the server. Since different technologies---even when just considering language workbenches only~\cite{Drux22}---handle composition in a different manner, the SPL client backend is the most likely portion of the architecture that needs to be re-implemented when changing technological space.

The \inlinebox[colback=red!10]{SPL server} receives the information regarding software assets from the client backend, generates the feature model and sends it to the client frontend.
In fact, the feature model generation algorithm is possibly the most important component of the SPL server: in this work we adopt a tag-based bottom-up feature model construction algorithm that was already tested in literature (see \sect{sect:contribution:fm}), but this component is intended to be fully customizable, as long as it is able to consume the same pieces of information.
Please notice that in this stage the pieces of information received by the server hold no notion regarding the fact that the original software assets were written in \tool{Neverlang}, instead it will rely on generic concepts such as artifacts, features, tags and atoms (more on that later).
Upon the user selecting a configuration, the client frontend and the server perform a back and forth communication apt at validating the configuration itself. Eventually, the user will commit a valid configuration that the server sends back to the client backend. Finally, as previously mentioned, the client backend translates the configuration into a product. In our example the product is a language interpreter or a compiler compatible with the \tool{Neverlang} runtime environment.

\subsection{Protocol}\label{sect:contribution:protocol}
The actors interact according to the following protocol; refer to the numbered steps in Fig.~\ref{fig:overview} for a visual overview.\smallskip

\noindent\textbf{The way forward.}\quad
The user develops the software assets \ding{182} that constitute the product line. The protocol may either be integrated into the development process from the outset or adopted retrospectively, allowing it to be applied to pre-existing software assets.
In our example, the user would develop \tool{Neverlang} modules akin to those shown in Listing~\ref{lst:loglangsrc} lines (1-13).

The user then interacts with the SPL client backend. Depending on the implementation, this backend may be a standalone command-line application or a component integrated directly into the development environment. Upon invocation, the SPL client backend inspects the software assets contained in the project \ding{183} and extracts the relevant information. Although the precise nature of this information will be detailed later in \sect{sect:contribution:components} and \sect{sect:contribution:messages}, it is sufficient for now to note that it describes the software artifacts and the capabilities they provide and/or require.
For instance, the \tool{Backup} module shown earlier would provide two artifacts: one for the syntax and one for the semantics.

The extracted information is subsequently transformed by the SPL client backend into a message whose representation is independent of the technological space of the original language assets \ding{184}. This representation can therefore be interpreted by the SPL server in terms of features and their dependencies.

The message is transmitted---either locally, via sockets or pipes, or over the network---to the SPL server. The server uses this message to generate a feature model \ding{185} that includes all features, their relationships and dependencies, as well as the mapping between features and the corresponding software artifacts. A possible output of this phase is a piece of metadata that represents a feature model akin to that shown in Fig.~\ref{fig:loglangfm}.
Notably, the feature model is unaware of the fact that the original artifacts were written in \tool{Neverlang}, thus the SPL server remains agnostic to both the development process of the artifacts and the specific technological space.

The feature model is subsequently serialized into a message and transmitted to the SPL client frontend, where it is rendered as an interactive graphical representation. Through this interface, the user can configure \ding{186} a product by selecting and deselecting features in the feature model.\smallskip

\noindent\textbf{The way back.}\quad
During the configuration process, each user interaction with the SPL client frontend triggers a message that is transmitted to the SPL server~\ding{187}. This architectural decision deliberately decouples the user interface from the configuration validation logic. In particular, the SPL client does not implement any mechanism for checking configuration validity or enforcing constraints. Instead, all validation logic is centralized on the server side, where consistency rules, feature constraints, and domain-specific requirements are evaluated in a uniform and authoritative manner. This centralization not only reduces redundancy but also ensures that validation behavior remains consistent across different client implementations. Consequently, the frontend can be replaced, extended, or customized to support alternative user experiences---such as graphical interfaces, command-line tools, or web-based configurators---without requiring any modification or reimplementation of the underlying validation logic.

In our example, we developed a simple yet complete graphical user interface that offers capabilities similar to other commonly used SPL engineering environments such as \tool{FeatureIDE}.

The interaction between user inputs and the SPL server progressively refines the configuration state and ultimately leads to the construction of a complete product configuration~\ding{188}.
In this process, the client and the server establish a contract in which the client trusts the server to perform all relevant checks regarding the validity of the configuration.
This configuration is therefore validated by the server and guaranteed to satisfy all structural and semantic constraints defined by the product line. Once validated, the configuration is considered ready as a concrete software product. At this stage, it is sent back to the SPL client backend, which leverages technological space–specific knowledge to translate the validated configuration into a final product~\ding{189}. In our case, the client backend would receive the list of all the active features from the server and use it to generate all the required slices (akin to Listing~\ref{lst:loglangsrc} lines 14-18) and a language unit (akin to Listing~\ref{lst:loglangsrc} lines 19-24).

Depending on the technological space and the specific needs of its users, the client backend can either execute the product directly or generate it as a piece of source code, that can be later compiled, distributed or even modified manually using traditional development workflows.
In our example, we opted to simply generate the slices and language source code to be later fed to the \tool{Neverlang} compiler.

\subsection{Feature Model Structure}\label{sect:contribution:fm}
In this section, we discuss the structure adopted for the technological space–agnostic feature model and its constituent features. The proposed design largely adheres to established practices in the literature, while incorporating certain specific choices that warrant further discussion.
Please notice that, due to the modularity of the proposed architecture, it is relatively easy to replace the feature model hereby presented and its generation algorithm with alternatives without affecting the rest of the architecture, namely the SPL client and the software assets, and, most importantly, the content of the messages sent among the actors.\smallskip

\noindent\textbf{Hierarchical Structure.}\quad
The feature model is organized according to a hierarchical structure---\textit{i.e.}, a tree---that serves as its structural backbone. Within this tree, some features represent alternative realizations of the same application concern, whereas others correspond to subcomponents of a broader functionality. This hierarchical organization provides a clear decomposition of concerns and supports a systematic representation of variability.

On top of this backbone tree, we superimpose a graph whose edges capture the dependencies among features. In other words, the tree encodes structural relationships, while the graph represents crosscutting constraints and inter-feature dependencies. Through this hybrid approach, we preserve a well-structured and comprehensible feature model while simultaneously enabling the explicit representation of all relevant dependencies.

In more complex scenarios, certain dependencies may be influenced by the values of feature attributes. These attributes are determined dynamically at runtime during the configuration process, according to user inputs. Such inputs may modify only the dependency edges in the graph, leaving the structural edges of the underlying tree unchanged. This separation between structural relationships and dynamic dependencies aims to keep the model stable and interpretable, even in the presence of runtime variability.\smallskip

\noindent\textbf{Tagging Software Components.}\quad
Features that relate to similar concerns are grouped by means of the notion of \textit{tag}, in a manner consistent with literature~\cite{Cazzola15f}. A tag is a label that provides a concise, high-level description of the concerns in which a feature is involved, while remaining sufficiently abstract to avoid being tied to a specific technological context. A feature may show multiple tags, suggesting that it is involved with several concerns.

Although the SPL server processes tags in a standardized and uniform way dictated by the feature model generation algorithm, the responsibility for assigning tags to software components lies with the SPL client backend. Depending on the chosen implementation strategy, tags may either be provided manually by the user or automatically inferred and attached by the backend.\footnote{Although feasible, writing protocol messages by-hand is not advised. If the technological space does not natively support any tagging mechanisms, one may use code comments with a special syntax that can be parsed by the SPL client backend.}
In our running example, we opted for the latter, as \tool{Neverlang} provides a tagging mechanism, as shown in Listing~\ref{lst:loglangsrc} line 5.
This design ensures flexibility while preserving a common semantic interpretation of tags at the server level.

Each feature may be associated with zero or more tags. For instance, consider a feature responsible for managing a motion sensor in a residential environment. Such a feature could be tagged with \textit{sensor}, \textit{motion}, and \textit{security}. The \textit{sensor} tag identifies its relation to sensing functionalities; \textit{motion} emphasizes its role in detecting movement; and \textit{security} highlights its potential use in intrusion monitoring. This tagging mechanism facilitates the grouping and management of related features, as features sharing one or more tags are likely to address overlapping or closely related concerns.\smallskip

\begin{algorithm}[t]
    \begin{algorithmic}[1]
    \State \textbf{begin}

        \State tagInstances(t) := length([ c for c in p.children() if t in c.tags() ])
        \State T = [ $t$ for $t$ in c.tags() for $c$ in p.children() if tagInstances(t)>1 ]
        \While{length($T$) > $0$}
            \State $t'$ = t with max tagInstances in T

            \State $n$ = newNode(name = $t'$)

            \For{\textbf{each} $c$ in $p$.children()}
                \If{t in c.tags()}
                    \State $c$.removeTag($t'$)
                    \State p.removeChild($c$)
                    \State n.addChild($c$)
                \EndIf
            \EndFor

            \State p.addChild(n)
            \State T = [ \textit{t} for \textit{t} in c.tags() for \textit{c} in p.children() if tagInstances(t)>1 ]
        \EndWhile
        \For{\textbf{each} $c$ in $p$.children}
            \State \textit{buildFeatureTree}($c$).
        \EndFor

    \State \textbf{end}
\end{algorithmic}
    \caption{buildFeatureTree(p:Node)}
    \label{lis:aideAlgorithm}
\end{algorithm}

\noindent\textbf{Feature Tree Derivation.}\quad
Tags are used to automatically derive the initial backbone of the feature tree from the set of software components by means of Algorithm~\ref{lis:aideAlgorithm}, which was part of a prior work~\cite{Cazzola15f}.
Recall that this is only one possible algorithm and other options may prove as effective.
The construction process begins with a dummy tree in which all features are direct children of a single, tagless root node. This initial structure serves as a neutral starting point for hierarchical refinement.

At each iteration, the most frequently occurring tag is identified and promoted to an abstract feature. All sibling nodes containing that tag are then relocated as children of the newly created abstract node, and the extracted tag is removed from those features. When no tag appears more than once among a set of siblings, the algorithm proceeds recursively on each child node.

Despite its conceptual simplicity, the algorithm exhibits several noteworthy properties. First, it is guaranteed to produce a tree structure. Since it is initially invoked on a tree and each transformation step either relocates, removes, or retains child nodes without introducing multiple parent assignments, no node can become a child of more than one parent. Consequently, the resulting structure is free from overlapping features.
Second, because the algorithm generates an elementary feature model, all nodes are optional with respect to their parent. This reflects the absence of mandatory constraints at the structural level and delegates additional constraints to the dependency graph.
Finally, the algorithm can accommodate crosscutting constraints by distributing them across different branches of the derived feature tree. In this way, structural organization and crosscutting relationships remain conceptually separated yet consistently represented within the overall model.

The final configuration of the \textit{feature model tree} comprises three types of nodes:

\begin{compactitem}
    \item \textit{root node}, which represents the principal element of the model. It must be activated in every valid configuration and represents the program as a whole.
    \item \textit{central nodes}, corresponding to abstract features introduced from tags originally associated with concrete features. These nodes serve a purely structural purpose and organize the feature space.
    \item \textit{leaf nodes}, which correspond to concrete features. Only these nodes may participate in dependency edges. Users are primarily expected to interact with these nodes during the configuration process.
\end{compactitem}

\subsection{Protocol Components}\label{sect:contribution:components}
We can now examine the information exchanged through the protocol for the purpose of extracting a feature model from linguistic assets in detail.\smallskip

\noindent\textbf{Artifacts.}\quad
We dub ``artifact'' any piece of code that can be part of one or more features in the product line. Depending on the technological space of choice, an artifact can either be a class, a function, or, in the context of language workbenches, a grammar production, a semantic asset or a combination of the two. Either way, the protocol is not concerned with how entity are implemented, as such choice is left to the implementation of the SPL client backend, but rather to the association between artifacts and features. In this context, the following statements are true:\smallskip
\begin{compactitem}
   \item a feature is comprised of one or more artifacts;
   \item an artifact is part of one or more features.\smallskip
\end{compactitem}
This leaves almost complete flexibility in terms of artifacts and their reusability: depending on the implementation of the SPL client backend, users can either opt for a one-to-one mapping between artifacts and features, let an artifact be part of only one feature or leave the complete flexibility of reusing an artifact across several features. For instance the \texttt{for} and \texttt{while} loops may share part of their semantics---\textit{i.e.}, some artifacts---while being different features.\smallskip

\noindent\textbf{Feature Dependencies.}\quad
Features inherently establish relationships with other features in order to accomplish specific functional objectives~\cite{Ye05}.
A feature dependency arises when one or more program elements---such as methods or attributes---belonging to a given feature rely on elements that reside outside the structural boundaries of that feature. A representative example is the definition of an attribute within one feature that is subsequently accessed or utilized by another feature~\cite{Cafeo16}. Such references introduce a level of coupling that must be explicitly represented in the feature model to ensure that the configuration process leads to a valid product.

In practice, we associate dependencies with software artifacts rather than directly with features. Then, features derive their dependencies from the artifacts they are comprised of---\textit{i.e.}, the dependency set of a feature is the union of the dependencies of its artifacts.
For instance, the \texttt{Backup} syntax presented in Listing~\ref{lst:loglangsrc} is, of course, incomplete: it lacks the definition on how to expand the \texttt{String} nonterminal, among other things; therefore it will depend on other syntactic assets that ``know'' how to perform this expansion. Conversely, it ``knows'' how to expand the \texttt{Cmd} nonterminal (on the left-hand side of the second production) and therefore it will be able to satisfy a dependency of any syntactic assets having \texttt{Cmd} on the right-hand side of a production.

From a configurational perspective, dependency relations impose constraints on feature activation. If a feature depends on another feature, the selection (activation) of the former logically entails the selection of the latter. Additionally, certain features may be mutually incompatible. For instance, when two features require exclusive access to a non-shareable resource, they cannot be simultaneously activated.\smallskip

\noindent\textbf{Atoms.}\quad
In order to represent the complete range of possible dependencies without making any assumptions on the technological space, we use the concept of an \textit{atom}. An atom is defined as a token that can either be required or provided by an artifact---and therefore by a feature. An atom is modeled as a dictionary-like structure, encoding arbitrary information.

Atoms constitute the minimal unit of dependency information within the protocol. An artifact may declare that it provides a given atom, requires a given atom, or both, depending on its functional role within a configuration.

Going back to the previous example, a very reasonable expression of the dependencies for the \texttt{Backup} syntax from Listing~\ref{lst:loglangsrc} in \tool{JSON} format is the following.
\showjson{atomBackupDependencies.json}


\noindent\textbf{Requirements Types.}\quad
You may have noticed the ``all'' key in the prior example. In fact, to represent the full range of possible relationships between features, we distinguish four distinct types of requirements:

\begin{compactitem}
    \item \textbf{all}---atoms that must be present for the feature to function correctly;
    \item \textbf{not}---atoms that are incompatible with the feature;
    \item \textbf{any}---groups of atoms in which at least one must be present;
    \item \textbf{one}---groups of atoms in which exactly one must be present.
\end{compactitem}

To capture more intricate dependencies, such as \textit{any}-type requirements between two groups of atoms where at least one group must be present, abstract features can be used.
These abstract features can require the individual atoms while simultaneously providing their combination, thereby expressing higher-level dependency structures.

For instance, a syntactic definition requires that ``all'' nonterminals on the right-hand side are provided by at least one feature, whereas semantic assets require that exactly ``one'' implementation of a data structure is available to work properly (such as, the \texttt{\$\$FileOp} data structure in Listing~\ref{lst:loglangsrc}, line 10).


\noindent\textbf{Variables and Globals.}\quad
To further enhance variability and compatibility across different contexts, some dependencies may be expressed in terms of the value of specific attributes that can be set at configuration time. Feature models supporting feature attributes are often referred to as ``extended feature models''.

To support this capability, we introduce the concepts of local and global variables, which will henceforth be referred to as \textit{variables} and \textit{globals}, respectively, for brevity.

\textit{Variables} are bound to a particular artifact implementing one or more features.
For example, consider a feature \texttt{display} that can operate in either a light or dark interface mode, implemented via an artifact \texttt{display concrete}. A variable \texttt{\$mode} can be used to specify the interface requirements of this artifact, as illustrated below.

\showjson{artifactDisplay.json}

\textit{Globals} adhere to the same principle, with the distinction that they are shared across the entire SPL\@. Any modification to a global variable affects all features within the system.
For instance, \tool{Neverlang} provides a language-wide nonterminal renaming mechanism (please refer to~\cite{Cazzola21b} for more details) that allows for any nonterminal to be renamed at will. This functionality was introduced to fix the incompatibility among language assets that were originally compatible except for the nonterminal used.\footnote{Consider how an asset for additive expressions with the syntax \texttt{Sum <-- Term "+" Expr} would be incompatible with another asset with the \texttt{Expr <-- AddExpr} syntax; the two assets can be rendered compatible by renaming \texttt{Sum} to \texttt{AddExpr}.}
In this context, all \tool{Neverlang} nonterminals can always be considered as globals whose value can be set across the entire configuration, therefore, if globals are highlighted with the \texttt{@} symbol, the dependencies of the \texttt{Backup} syntax can be rewritten as follows.

\showjson{variablesBackupDependencies.json}

If a required or provided atom is defined in terms of a variable or a global, then changing its value may affect the dependencies between features, because a specific dependency may be satisfied only if specific attributes have a specific value at configuration time.

\subsection{Messages}\label{sect:contribution:messages}
We can finally introduce the set of messages that are sent across the protocol's components to convey all necessary information with the necessary degree of generality. As shown in the examples provided so far, we opted for a \tool{JSON} implementation of messages in this work, although other options are of course viable. Although examples are omitted for brevity, please consider that all messages share this structure. We also assume that the connection is reliable, thus omitting any acknowledgment messages.\smallskip

\noindent\textbf{\large\texttt{features}.}\quad
Used to load the SPL.\\
\indent\texttt{Content}: Collection of artifacts and features within the workspace, each with its list of composing artifacts. Each artifact declares its own dependencies, tags and attributes. Features can require only artifacts among this collection.\\
\indent\texttt{Response}: None.\\
\indent\texttt{Usage}: The client backend sends the currently available language assets to the server to trigger the generation of the feature model according to Algorithm~\ref{lis:aideAlgorithm}.\smallskip


\noindent\textbf{\large\texttt{featureModel}.}\quad
Used to display the feature model in the client frontend.\\
\indent\texttt{Content}: The feature model generated by the server, comprised of the features and their relations.\\
\indent\texttt{Response}: None.\\
\indent\texttt{Usage}: Upon computing the feature model, the SPL server sends it to the SPL client backend for it to be displayed to the user.\smallskip

\noindent\textbf{\large\texttt{updateAttribute}.}\quad
Used to change the value of an attribute.\\
\indent\texttt{Content}: Feature name, attribute name, value. Together, these pieces of information allow unique identification of an attribute within the feature model.\\
\indent\texttt{Response}: Any change to the feature dependencies within the feature model.\\
\indent\texttt{Usage}: Notifies the SPL server about the fact that the user changed the value of a feature attribute. Such a change may trigger updates to feature dependencies. The SPL server computes any updates and sends it back to the SPL client backend.\smallskip

\noindent\textbf{\large\texttt{activate}.}\quad
Used to toggle the activation status of a feature in the current configuration.\\
\indent\texttt{Content}: Name of the feature to activate or deactivate.\\
\indent\texttt{Response}: List of currently active features.\\
\indent\texttt{Usage}: Notifies the SPL server about the fact that the user changed the activation status of a feature. Such a notification causes the server to updated the internally stored configuration, updating the set of active features. This does not only include the selected feature, but also any dependent features. The resulting set is sent back to the SPL client frontend to updated the view presented to the user.\smallskip

\noindent\textbf{\large\texttt{validate}.}\quad
Used to check if a configuration is valid and to obtain hints on how to fix it.\\
\indent\texttt{Content}: None.\\
\indent\texttt{Response}: A map whose keys are the names of any invalid features and whose values are the respective unfulfilled requirements.
These requirements are expressed through a dictionary with four possible keys: \texttt{ALL}, \texttt{NOT}, \texttt{ANY}, \texttt{ONE}, each mapped to the atoms required but not provided or vice-versa.
The response also contains providers for each atom---\textit{i.e.}, the list of features that could resolve the dependency by being activated or deactivated.\\
\indent\texttt{Usage}: request the server to check if the requirements for each active features are fulfilled. If the response is positive, the configuration is valid and can be committed, otherwise the server responds with hints on how to reach a valid configuration.\smallskip

\noindent\textbf{\large\texttt{commit}.}\quad
Used to export the current configuration from the SPL client frontend all the way back to the language assets.\\
\indent\texttt{Content}: List of active features and attribute values.\\
\indent\texttt{Response}: None.\\
\indent\texttt{Usage}: Notifies the SPL server that the user wants to translate the current valid configuration into a product. The same message is then forwarded from the SPL server to the SPL client backend, triggering the product generation. How this is performed depends on the chosen language workbench and the SPL client backend implementation. Regardless, the expected behavior is for the SPL client backend to generate a new language and add it to the workspace among other language assets.

\section{Case Study}\label{sect:casestudy}

We briefly present a prototypical implementation of an architecture orchestrated according to the protocol. An in-depth discussion of the implementation details is beyond the scope of this paper, although this case study serves the purpose of demonstrating the applicability of our contribution.
Each paragraph focuses on a component that can be independently developed and replaced to adapt to different requirements and scenarios.
\smallskip

\noindent\textbf{SPL client backend.}\quad
The SPL client backend is comprised of two \tool{Neverlang}-specific adapters.
The first adapter generates the \textit{featureModel} message in \tool{JSON} format by extracting all relevant pieces of information from the \tool{.class} files obtained through the compilation of \tool{Neverlang} source files.
The client backend is implemented as a \tool{Java} library that creates an instance of available language assets and then uses \tool{Neverlang}'s reflective capabilities~\cite{Cazzola22d} to create the appropriate artifacts and features for each module.
Dependencies and the respective atoms used by the protocol are determined on a syntactical level by considering required and provided nonterminals: each module's productions are analyzed to identify nonterminals that are not provided by the module itself and that must be provided elsewhere.
Each nonterminal is associated with an attribute that reflects its name, allowing the user to rename it and modify the corresponding dependencies possibly filling any gap existing within the grammar~\cite{Cazzola20}.

Regarding the semantics, the client creates a feature for each semantic role---\textit{i.e.}, visit of the parse tree---each with two available attributes: priority and visiting order.

The second adapter takes the validated configuration from the SPL server and generates a \tool{Neverlang} language, akin to Listing~\ref{lst:loglangsrc} lines 19-24. This adapter does not require any reflective capabilities, as it simply translates a \tool{JSON} input into \tool{Neverlang} syntax, then writes the result to file.\smallskip

\noindent\textbf{SPL server (\tool{Go}).}\quad
The server core implementation leverages the \texttt{"net/http"} standard \tool{Go} library package. 
Most relevant information, including feature and artifact names, tags, attribute names and values, hashes, and even atoms can all be represented as strings. To avoid any ambiguous data structures and leverage the strong typing guarantees,
we introduced various alias types for the string type.

Local and global attributes are represented as map data structures.

The natural choice for implementing atoms is using dictionaries, as the \texttt{"encoding/json"} package deserializes information into a map by default. 

Throughout the implementation, the server stores hashes rather than strings whenever possible to ensure that their format is compatible with the atom naming conventions needed by the \tool{Prolog} validator.

Notably, the \tool{Go} core also implements the feature model generation algorithm. This portion of the code is meant to be replaced if the proposed algorithm is unsuitable for the needs of the specific technological space.\smallskip

\noindent\textbf{SPL server (\tool{Prolog}).}\quad
While \tool{Go} offers a vast standard library to implement client-server communication and serialization capabilities, we deemed a logic programming language as \tool{Prolog} more suitable to implement the configuration validation capabilities offered by the server.

\begin{listing}[t]
    \centering
    \showprolog*{exists.pl}
    \caption{Atom existence property.}\label{lis:exists}
\end{listing}

\begin{listing}[t]
    \centering
    \showprolog*{valid.pl}
    \caption{Feature validity property.}\label{lis:valid}
\end{listing}

\begin{listing}[t]
    \centering
    \showprolog*{dependencies.pl}
    \caption{Requirements validity checks implementation.}\label{lst:requirements}
\end{listing}

As shown in \sect{sect:background}, \tool{Prolog} programs are built from \textit{atoms}---the base components of logical statements used in declarations and relationships.
Note that this use of ``atom'' is specific to \tool{Prolog} terminology and distinct from the protocol-level atoms introduced in \sect{sect:contribution}.
The validation process we chose to implement is based on two fundamental principles: atoms may exist (Listing~\ref{lis:exists}), and features may be valid (Listing~\ref{lis:valid}).
Intuitively, an atom exists when it is provided by a valid feature, whereas a feature is considered valid when all the following conditions are met:
\begin{compactitem}
    \item it provides at least one atom,
    \item all of its required atoms are provided by some other feature,
    \item none of its undesired atoms are provided by any feature.
\end{compactitem}
Verifying the existence of an atom is straightforward, whereas determining a feature's validity is more complex, as it requires assessing all dependencies types while preventing circular dependencies.\ \tool{Prolog}'s expressive capabilities enable a relatively simple implementation of these dependency checks. There are five required checks:
\begin{compactitem}
   \item \inlineprolog{provides(feature, atom)}, 
   \item \inlineprolog{requisitesAll(feature, atom)}, 
   \item \inlineprolog{requisitesNot(feature, atom)} 
   \item \inlineprolog{requisitesAny(feature, atom, groupID)}, and 
   \item \inlineprolog{requisitesOne(feature, atom, groupID)}. 
\end{compactitem}

The logic implemented in \tool{Prolog} goes beyond checking configuration validity.
As previously mentioned, the validation process also provides suggestions regarding which features to activate or deactivate to adjust it.
To achieve this, the system first evaluates the validity of all active features and records those that are found to be invalid with their missing requirements.
Then, for each missing requirement of each invalid feature it finds any possible providers and reports them to the user through the response to the \texttt{validate} message.

In summary, the \tool{Prolog} dependency validation logic represents the core of the SPL server as it serves both the purposes of validating a configuration and of informing the user on how to fix an invalid configuration. Despite that, thanks to the capabilities offered by \tool{Prolog} in terms of querying the knowledge base, the implementation itself is rather succinct, as it amounts to a total of only 32 lines of code, as shown in Listing~\ref{lst:requirements}.
\smallskip

\noindent\textbf{SPL client frontend.}\quad
To demonstrate the interaction with the protocol from a user's perspective, we developed a simple graphical user interface displaying changes in the feature model graph.
The interactive graph is implemented through a versatile \tool{Javascript} library called \tool{Cytoscape}~\cite{Shannon03}.
To provide an effective means of interacting with feature attributes, we integrated another library called \tool{Tippy}\footnote{\url{https://atomiks.github.io/tippyjs/}} used to create interactive pop-ups on the nodes and the edges of the graph.

Please consider that we made the deliberate choice of not using any well-known variability management software such as \tool{FeatureIDE} to avoid introducing implementation bias, although in real-world use cases it may be beneficial to implement the SPL client frontend as a plugin for existing software rather than from scratch.

Our prototype can show all relevant kinds of information, including feature nodes (both concrete and abstract), structural edges (parent-child) and dependency edges, dead features and so on.
Each interaction with the graph, be it feature activation or deactivation, attribute value edit, or configuration commit, causes the corresponding message to be sent from the client to the server. In case of an \textit{updateAttribute} message, the response may cause the feature model to update; in this case, the graphical representation presented by the client to the user is updated accordingly.\smallskip

\begin{figure*}[t]
    \mbox{\hspace{-.5cm}%
   \begin{subfigure}{.55\textwidth}
      \centering
         \includegraphics[width=.92\textwidth]{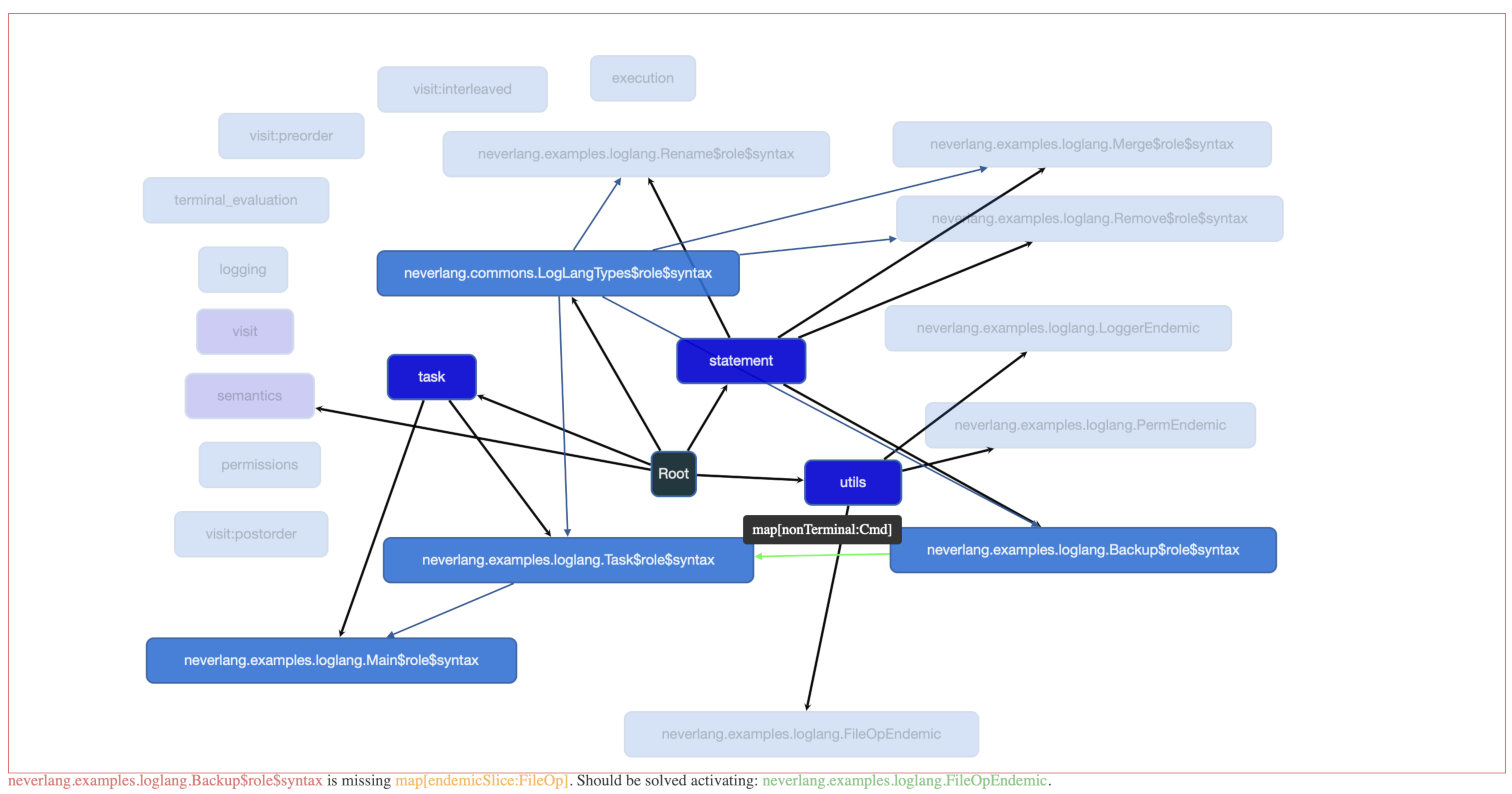}
         \caption{Feature selection.}\label{fig:configProblemOpen}
     \end{subfigure}\hspace{-1cm}
   \begin{subfigure}{.55\textwidth}
       \centering
         \includegraphics[width=.92\textwidth]{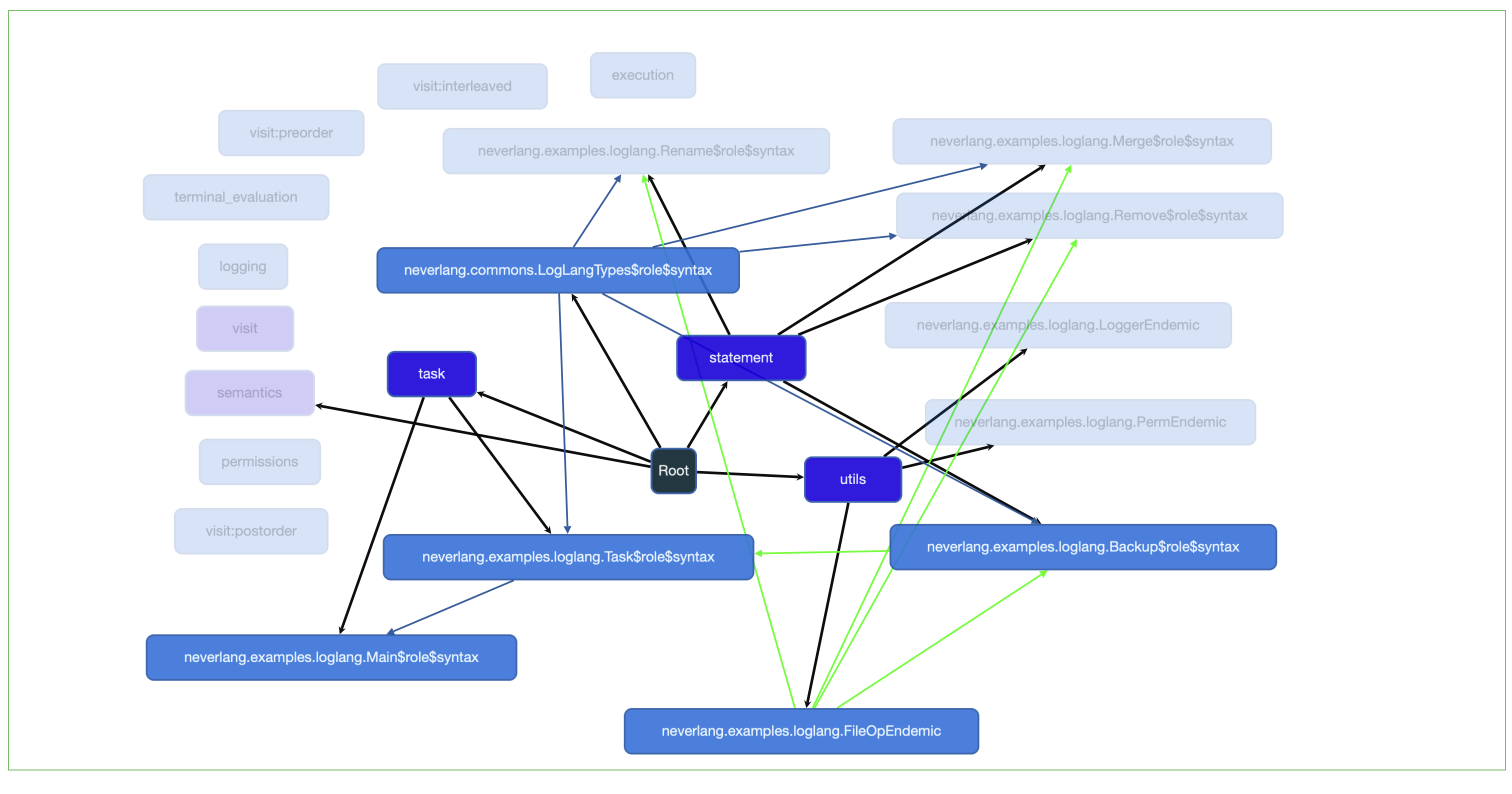}
         \caption{Dependency resolution.}\label{fig:configProblemSolved}
   \end{subfigure}}
   \caption{Exemplary configuration process step within the SPL client frontend.}\label{fig:configProblem}
\end{figure*}

\noindent\textbf{Configuration process.}\quad
The protocol implementation was tested against a \tool{LogLang} and a \tool{Javascript} implementation written in \tool{Neverlang}, the latter comprised of 50 programming concepts implemented in 73 \tool{Neverlang} slices. Through the SPL client frontend, we were able to configure all aspects of a \tool{Javascript} dialect, starting from the configuration process, all the way back to the language generation and usage, including intermediate steps such as nonterminal renaming, syntax and semantics configuration.
We replicated experiments we performed in prior works, such as the gradual extension of \tool{Javascript} for didactic purposes~\cite{Cazzola21b}.
However, the frontend is still in a proof-of-concept state and its usability with large LPLs such as \tool{Javascript} is limited. Thus, we will hereby showcase configuration examples regarding \tool{LogLang}:
the client frontend and the configuration process are inspired by products such as \tool{FeatureIDE}, whose description is beyond the goal of this paper, but please let us showcase the intended workflow and summarize the capabilities of the protocol with an example.
First, the user invokes the client backend that inspects the \tool{LogLang} software assets and generates a collection of artifacts, features and their dependencies in \tool{JSON} format, according to the structure presented in \sect{sect:contribution}. Then, this \tool{JSON} is parsed by the SPL server in one of three possible modes:
\begin{compactitem}
    \item read from pipe between client and server;
    \item loaded from file through the command line;
    \item loaded from file through GUI.
\end{compactitem}
This triggers the feature model generation algorithm and the result is forwarded to the client frontend that graphically presents it to the user.

Consider the case presented in Fig.~\ref{fig:configProblem}, showing a screenshot of the client frontend in which the \tool{LogLang} feature model was loaded and in which the user is manually configuring a product.
Fig.~\ref{fig:configProblemOpen} contains a partial (invalid) configuration that cannot be committed; as suggested in the box---highlighted in red due to the configuration being invalid---, the dependencies can be solved by activating the \tool{FileOpEndemic} feature.\footnote{Package names and role names omitted for brevity.}
Fig.~\ref{fig:configProblemSolved} shows the same configuration process after activating the suggested feature; as a result, the box is highlighted in green: no more features need to be activated or deactivated and thus the configuration is valid and can be committed.

Other configuration aspects were tested in a similar manner; most notably, configuring the semantics, eventually led to a selection among one of three possible features: \texttt{visit:preorder}, \texttt{visit:postorder}, and \texttt{visit:interleaved} (currently not selected and therefore greyed out on the left side of Fig.~\ref{fig:configProblemOpen}). The user can decide the parse tree traversal strategy for the language semantics by activating the corresponding feature.
As dictated by the SPL client backend, each feature concerned with a semantic role has a ``priority'' local variable that accepts an integer value. Upon committing a valid configuration, the value of this variable will be used by the client backend to determine the role execution order (each role representing a parse tree visit).

The configuration process may eventually lead to a situation in which the partial configuration cannot be fixed by taking any additional actions. In this case, the box will be red as in Fig.~\ref{fig:configProblemOpen}, but there will be no available suggestions on the bottom, since the validation step performed by the \tool{Prolog} server was unable to find any features whose provided atoms could render the configuration valid.

Please note that this front-end application is agnostic to both the language workbench used and to the server-side validation process: it simply displays the information received from the server and sends back any inputs received from the user. This means that the same experiment could be replicated using a different frontend, provided it can send/receive the correct messages from/to the server.\smallskip

\noindent\textbf{Implementation effort.}\quad
Due to the qualitative nature of our contribution, a full-fledged empirical evaluation is beyond the scope of this paper, however let us complete this section by briefly estimating the development effort needed to implement each component in the architecture.

The implementation is in a proof-of-concept state that could be rendered more complex with more sophisticated functionalities and because implementing different components using different languages may present some trade-offs. For instance, swapping the \tool{Prolog} core with a different language may offer better performance at the cost of a more verbose implementation.
The generality of such a discussion is therefore questionable, but we hope readers will be able to perceive the potential benefits of using such an architecture to extract product lines in different technological spaces.

Our implementation is comprised of 2551 lines of code (LoC) in total, of which:
\begin{compactitem}
    \item 1339 LoC of SPL server (\tool{Prolog} + \tool{Go});
    \item 535 LoC of SPL client frontend (\tool{Javascript});
    \item 677 LoC of SPL client backend (\tool{Java}).
\end{compactitem}

\noindent We can further divide the SPL server into the following:
\begin{compactitem}
    \item 53 LoC of \tool{Prolog} core;
    \item 410 LoC of feature model generation algorithm;
    \item 876 LoC of other server functionalities (message generation, parsing etc.).
\end{compactitem}

\noindent Each of these five components---Prolog validation core, feature model generation algorithm, Go server infrastructure, Java client backend, and JavaScript client frontend---can be reused independently.
In summary, this gives us a rough estimate of the potential reduction in implementation effort in the following scenarios:
\begin{compactenum}
    \item 84\% less effort in replacing the feature model generation algorithm without changing the technological space nor the rest of the server nor the client;
    \item 98\% less effort in swapping the validation logic without changing the technological space, nor the rest of the server nor the client;
    \item 48\% less effort in replacing the entire server without changing the rest;
    \item 73\% less effort in using the architecture with a different technological space (for instance using a different language workbench instead of \tool{Neverlang}, thus replacing the \tool{Java} backend) while reusing the rest of the architecture;
    \item 79\% less effort in replacing the client frontend (for instance, using a different GUI or an existing tool such as \tool{FeatureIDE}) while reusing the rest of the architecture.
\end{compactenum}

\noindent Among these cases, we consider case number 4 to be the most likely, since the components representation and the composition mechanisms can differ wildly depending on the technological space, whereas the other cases are a matter of preference with regards to the extraction and configuration logic and of user experience. Of course, the actual implementation effort may vary depending on the actual technology being used and increases if more than one component needs to be replaced at the same time, however the modular architecture allows developers to tune the effort to their needs.

\section{Related Work}\label{sect:related-work}
This section positions our contribution with respect to feature model extraction, language workbenches, protocol-based approaches, feature model analysis, and compositional language development.

\subsection{Feature Model Extraction and SPL Adoption}
Our work follows the bottom-up, extractive approach to SPL engineering originally identified by Krueger~\cite{Krueger01}.
Several techniques have been proposed for reverse-engineering feature models from existing assets.
She~\textit{et~al.}~\cite{She11} reverse-engineer feature models from natural-language descriptions of configuration options (\textit{e.g.}, \tool{Kconfig}), whereas our protocol operates on generic artifacts and their provide/require relationships without assuming a particular variability encoding.
Acher~\textit{et~al.}~\cite{Acher14} focus on extracting, merging, and slicing feature models from heterogeneous artifact descriptions, treating feature models as first-class entities---complementary to our emphasis on a workbench-agnostic extraction protocol.
Haslinger~\textit{et~al.}~\cite{Haslinger13} use formal concept analysis to extract feature models from source code; unlike their implementation-level focus, our protocol abstracts over artifact representations through atoms, making it applicable across technological spaces.
Martinez~\textit{et~al.}~\cite{Martinez15} developed \tool{BUT4Reuse}, a bottom-up framework for extracting reusable elements from product variants. While \tool{BUT4Reuse} identifies commonalities and variabilities across existing products, our protocol extracts a feature model from modular artifacts organized around explicit dependency declarations.
Assun\c{c}\~{a}o~\textit{et~al.}~\cite{Assuncao17} survey re-engineering approaches for SPL adoption; our contribution extends this landscape with a protocol-level abstraction that decouples the extraction process from both the language workbench and the development environment.

\subsection{Language Workbenches and Modular Languages}

Language workbenches provide integrated environments for defining and composing programming languages, and several support modular development to varying degrees. Our contribution is orthogonal to the choice of workbench and aims at bridging different technological spaces through a common protocol.

\tool{Spoofax}~\cite{Wachsmuth14,Kats10} relies on declarative meta-languages for syntax, transformations, and type checking, but its tooling is tightly coupled to its own meta-languages and to the \tool{Eclipse} platform.
\tool{MPS}~\cite{Voelter12b} adopts a projectional editing approach that supports language composition via extension and embedding, yet its tight IDE integration makes it difficult to decouple the SPL engineering process from the platform.
\tool{Xtext}~\cite{Bettini16} generates parsers and editors from grammar specifications within \tool{Eclipse}, offering grammar inheritance and mixin composition but limited SPL support.
\tool{MontiCore}~\cite{Krahn10} promotes grammar-based modular language development through inheritance, embedding, and aggregation, yet its feature modeling capabilities remain tied to its own infrastructure.
\tool{Rascal}~\cite{Klint09} offers powerful facilities for source code analysis and language prototyping but lacks dedicated SPL engineering support.
\tool{Melange}~\cite{Degueule15} enables assembly of reusable language modules through merge, slice, and inherit operators, though its composition mechanisms are specific to the \tool{EMF} technological space.
\tool{SCOLAR}~\cite{Pfeiffer21, Pfeiffer23b} supports language aggregation of black-box components in a low-code platform that follows principles similar to those adopted by product line engineering technologies.
\tool{LionWeb}\footnote{\url{https://lionweb.io/}} is an ecosystem to develop language-oriented systems on the web that handles clients access to functionalities through \textit{models} (graph data structures) and \textit{repositories}. The \tool{LionWeb} models are designed to interoperate and to reuse existing other pre-existing compatible components.

In all cases, SPL engineering capabilities---when present---are embedded within the workbench and cannot be reused across platforms. Our protocol addresses this by externalizing feature model extraction and configuration logic into an independent server communicating with workbench-specific backends through a standardized message format.

\subsection{Protocol-Based Approaches in SW Engineering}

Our protocol draws inspiration from the LSP, which decouples language-specific intelligence from editor concerns via a client-server architecture~\cite{Bunder19}. The Debug Adapter Protocol (DAP)\footnote{\url{https://microsoft.github.io/debug-adapter-protocol/}} and Build Server Protocol (BSP)\footnote{\url{https://build-server-protocol.github.io/}} apply the same principle to debugging and build tooling, respectively.

Bruzzone~\textit{et~al.}~\cite{Cazzola25b} applied this decoupling philosophy to language server generation for language families built with \tool{Neverlang}, proposing a modular generation process paired with an automated LSP plugin generator for multiple editors. They also introduce the \textit{variant-oriented programming} paradigm and a cross-artifact coordination layer for managing interdependent software variants. While their approach targets editing support (\textit{i.e.}, language servers and plugins), our protocol targets a complementary concern: feature model extraction, construction, and interactive configuration. Both could be combined for end-to-end, workbench-agnostic LPL tooling. To the best of our knowledge, no prior work other than ours has proposed a protocol-based decoupling specifically for SPL or LPL feature model extraction.

\subsection{Feature Model Analysis and Configuration}

Benavides~\textit{et~al.}~\cite{Benavides10} survey automated analysis operations on feature models, including detection of dead features, false-optional features, and void models. Our protocol delegates all validation logic to the SPL server, allowing any such technique to be integrated within its implementation.

Batory~\cite{Batory05} established a formal link between feature models, context-free grammars, and propositional formulas, enabling SAT-based~\cite{Mendonca09} and BDD-based~\cite{Czarnecki07} analysis and configuration. Our use of \tool{Prolog} as the reasoning engine offers an alternative that naturally supports constraint propagation and interactive exploration, while remaining general enough to accommodate other solvers. Czarnecki and Wasowski~\cite{Czarnecki07} additionally investigated feature model merging and specialization; although our protocol currently focuses on single-model extraction, the message-based architecture could support composition operations in future work.

Bruzzone~\textit{et~al.}~\cite{Bruzzone26-preprint} address a complementary downstream challenge: prioritizing configurations when their combinatorial explosion makes exhaustive exploration infeasible. They rank features via centrality measures on graph structures extracted from the \tool{Rust} compiler, with validity guaranteed by a SAT solver over a CNF encoding. While their work focuses on selecting relevant configurations for analysis and testing, our protocol targets the upstream extraction and interactive configuration of feature models; the two could be combined to guide configuration-aware optimization.

\subsection{Feature-Oriented and Compositional Language Development}

Feature-oriented software development~(FOSD)~\cite{Apel13} treats features as first-class modular units. Tools such as \tool{AHEAD}~\cite{Batory04} and \tool{FeatureHouse}~\cite{Apel09} support composition of feature modules across multiple artifact types; our protocol complements them by providing a standardized mechanism for extracting the variability structure governing such compositions.

Delta-oriented programming~\cite{Schaefer10} offers an alternative paradigm in which a core product is incrementally modified through delta modules, as in \tool{DeltaJ}. While delta-oriented programming focuses on how artifacts are composed, our protocol focuses on how their dependencies are represented, extracted, and validated in a workbench-agnostic manner.
More recently, \textit{variant-oriented programming}~\cite{Cazzola25b} has been proposed for managing interdependent software variants arising when language artifacts are reused across a language family. Through a cross-artifact coordination layer it ensures variant consistency---complementary to our protocol, which governs how variability is extracted and configured at the feature model level.

\tool{FeatureIDE}~\cite{Thum14} integrates feature modeling, configuration, and code generation within \tool{Eclipse}; \tool{AiDE}~\cite{Cazzola20} builds on it for LPL engineering with \tool{Neverlang}. Our protocol generalizes the role of \tool{FeatureIDE} by decoupling feature model management from both IDE and language workbench, enabling heterogeneous environments to interoperate through a shared protocol.

\section{Conclusion}\label{sect:conclusion}
We have presented a generalized, workbench-agnostic protocol for bottom-up
feature model extraction, with a particular focus on Language Product Lines.
The protocol abstracts from specific language workbenches and IDEs by
representing variability through atoms and their \texttt{provides}/\texttt{requires}
relations, and by clearly separating concerns between SPL server, backend, and
client.

We realized this protocol in a prototype that extracts a feature model from
\tool{Neverlang} language artifacts and exposes it through a
\tool{Cytoscape}-based graphical frontend. The case study illustrates that the
same SPL infrastructure can be reused across different technological spaces,
while backends remain thin adapters over existing artifacts.

Our current implementation targets a single workbench and a limited set of
artifacts; broader validation across additional workbenches and technological
spaces is left for future work, together with the integration of richer
analysis operations and configuration support. We envision our protocol as
complementary to protocol-based language tooling such as LSP, and plan to
explore their combination for end-to-end, workbench-agnostic LPL support.

\bibliographystyle{ACM-Reference-Format}
\bibliography{local,strings,logic,programming,software_engineering,dsl,pl,tools,splc,oolanguages,my_work,ml+nn,grammars,security,roles,cop,aosd,pattern,dsu,reflection}
\end{document}